\documentclass[12pt]{article}
\usepackage{epsfig,rotating,setspace,latexsym,amsmath,epsf,amssymb,bm}
\usepackage{cite}

\newcommand{\tr}{\text{tr}}

\newtheorem{theorem}{Theorem}

\newtheorem{lemma}{Lemma}

\newenvironment{proof}[1]{\medskip\par\noindent
{\bf Proof:\,}\,#1}{{\mbox{\,$\blacksquare$}\par}}

\newcommand{\av}{\mathbf{a}}

\newcommand{\gv}{\mathbf{g}}

\newcommand{\qv}{\mathbf{q}}

\newcommand{\Amat}{\mathbf{A}}

\newcommand{\Cmat}{\mathbf{C}}
\newcommand{\Dmat}{\mathbf{D}}

\newcommand{\Hmat}{\mathbf{H}}
\newcommand{\Imat}{\mathbf{I}}
\newcommand{\Nmat}{\mathbf{N}}

\newcommand{\Smat}{\mathbf{S}}

\begin{document}

\title{Towards the Secrecy Capacity of the Gaussian MIMO Wire-tap Channel:
The 2-2-1 Channel\thanks{This work was
supported by NSF Grants CCR $03$-$11311$, CCF $04$-$47613$ and CCF
$05$-$14846$.}}

%\author{b}
\author{Shabnam Shafiee \qquad Nan Liu \qquad Sennur Ulukus \\
\normalsize Department of Electrical and Computer Engineering \\
\normalsize University of Maryland, College Park, MD 20742 \\
\normalsize {\it sshafiee@umd.edu} \qquad {\it nkancy@umd.edu} \qquad {\it ulukus@umd.edu} }

\maketitle

\begin{abstract}
We find the secrecy capacity of the 2-2-1 Gaussian MIMO wire-tap
channel, which consists of a transmitter and a receiver with two
antennas each, and an eavesdropper with a single antenna. We
determine the secrecy capacity of this channel by proposing an
achievable scheme and then developing a tight upper bound that meets
the proposed achievable secrecy rate. We show that, for this channel,
Gaussian signalling in the form of beam-forming is optimal, and no
pre-processing of information is necessary. 
\end{abstract}

%\vspace*{0.4cm}\noindent {\em keywords:}
%%\begin{keywords}
%Wire-tap channel, multiple antennas, MIMO, information theoretic 
%security, secrecy capacity.
%%\end{keywords}

\newpage
%================================================================
\section{Introduction}

The inherent openness of wireless communications makes it vulnerable
to eavesdropping and jamming attacks. This vulnerability has to be
addressed through secure communications. The eavesdropping attack was
first studied by Wyner in \cite{Wyner:1975}, where he considers a
single-user wire-tap channel. The measure of secrecy is the message
equivocation rate at the wire-tapper, which is defined as the entropy
of the message at the wire-tapper, given the wire-tapper's
observation. Wyner models the wire-tapper's channel as a degraded
version of the channel from the transmitter to the legitimate
receiver, which is a reasonable assumption in a wired channel. For
this channel, Wyner identifies the rate-equivocation region and
therefore, the secrecy capacity. Wyner's result was extended to the
Gaussian wire-tap channel in \cite{Hellman:1978}, and it was shown
that Gaussian signalling is optimal. The secrecy capacity was found to
be the difference between the capacities of the main and the
eavesdropping channels.

Csiszar and Korner \cite{Csiszar:1978} studied the general, i.e., not
necessarily degraded, single-transmit-\break ter, single-receiver,
single-eavesdropper, discrete memoryless channel with secrecy
constraints, and found an expression for the secrecy capacity, in the
form of the maximization of the difference between two mutual
informations involving an auxiliary random variable. The auxiliary
random variable is interpreted as performing pre-processing on the
information. The explicit calculation of the secrecy capacity for a
given channel requires the solution of this maximization problem in
terms of the joint distribution of the auxiliary random variable and
the channel input. 
%Solving this optimization problem directly is
%difficult, forcing researchers typically to follow a two-step
%solution, where in the first step a feasible solution is identified
%(an achievable scheme), and in the second step a tight upper bound
%that meets this feasible solution is developed (tight converse).

%The use of multiple transmit and receive antennas has been shown to
%increase the achievable rates when there are no secrecy constraints
%\cite{Telatar:1999}. The Gaussian multiple-input multiple-output
%(MIMO) wire-tap channel is a special case of the single-transmitter,
%single-receiver, single-eavesdropper wire-tap channel, therefore,
%finding its secrecy capacity involves identifying the optimum joint
%distribution of the auxiliary random variable representing
%pre-processing and the channel input in the Csiszar-Korner formula.
%Since the Gaussian MIMO channel is not degraded in general, finding
%its secrecy capacity is challenging, just as it has been to find its
%information capacity without secrecy constraints
%\cite{Viswanath:2003,Vishwanath:2003,Yu:BC2004,Jindal:2005,Shamai:2006}.

The use of multiple transmit and receive antennas has been shown to
increase the achievable rates when there are no secrecy constraints
\cite{Telatar:1999}. The Gaussian multiple-input multiple-output
(MIMO) wire-tap channel is a special case of the single-transmitter,
single-receiver, single-eavesdropper wire-tap channel. Since the Gaussian MIMO channel is not degraded in general,
finding its secrecy capacity involves identifying the optimum joint
distribution of the auxiliary random variable representing
pre-processing and the channel input in the Csiszar-Korner formula. However, solving this optimization problem directly for non-degraded channels is
difficult, forcing researchers typically to follow a two-step
solution, where in the first step a feasible solution is identified
(an achievable scheme), and in the second step a tight upper bound
that meets this feasible solution is developed (tight converse).

The first paper studying secrecy in MIMO communications is
\cite{Negi:2005}, which proposes an achievable scheme, where the
transmitter uses its multiple transmit antennas to transmit only in
the null space of the eavesdropper's channel, thereby preventing any
eavesdropping. Reference \cite{Parada:2005} studies the Gaussian
single-input multiple-output (SIMO) wire-tap channel, and shows that
it is equivalent to a scalar Gaussian channel, and gives the secrecy
capacity using the results of \cite{Hellman:1978}. An achievable
scheme has been proposed for the Gaussian multiple-input single-output
(MISO) wire-tap channel in \cite{Shafiee:2007}, and independently and
concurrently in \cite{Li:2007}. In both of these papers, the
achievable secrecy rate is obtained by restricting the channel input
to be Gaussian, with no pre-processing of information. The secrecy
rate found in \cite{Shafiee:2007, Li:2007} is shown to be the secrecy
capacity of the Gaussian MISO wire-tap channel in
\cite{Khisti:ISIT,Khisti:MISOME}. 
%Inner and outer bounds on the secrecy capacity
%region for the Gaussian MISO broadcast channel are derived in
%\cite{LiuPoor:2007}, where two receivers wish to have simultaneous
%secret communications with the transmitter which has multiple
%antennas. 
Further, \cite{Khisti:ISIT,Khisti:MISOME} allow the eavesdropper to have multiple antennas (MISOME).

In all of the above papers, the secrecy capacity of MIMO communications is specified
only in the cases where the receiver has a single antenna. The next
step towards finding the secrecy capacity of the general Gaussian MIMO
channel is to consider multiple antennas at the receiver. In this
paper, we consider a MIMO channel where both the transmitter and the
receiver have multiple antennas. Since the general problem seems to be
intractable for now, we focus on a simple special case where both the
transmitter and the receiver have two antennas each, and the
eavesdropper has a single antenna, hence we call this channel the
2-2-1 MIMO wire-tap channel. We find the secrecy capacity in two
steps: we first propose an achievable scheme, which is a Gaussian
signalling scheme with no pre-processing of information, and then, we
develop a tight upper bound that meets the rate achieved with our
proposed signalling scheme. 

We first show that the optimal Gaussian signalling scheme has a
unit-rank transmit covariance matrix, hence with Gaussian signalling,
beam-forming is optimal. The transmitter beam-forms in a direction
that is as orthogonal to the direction of the eavesdropper, and as
close to the two directions of the receiver as possible. Then, we
develop an upper bound by considering a channel where the
eavesdropper's signal is given to the receiver. The secrecy capacity
of this channel is an upper bound to the secrecy capacity of the
original channel. In addition, this channel is degraded, and no
pre-processing of information is needed.  Furthermore, Gaussian
signalling is optimal for this channel. We further tighten this bound
by allowing correlation between the additive noises of the receiver
and the eavesdropper. For a certain such correlation, we prove that
the optimal Gaussian signalling is unit-rank in this upper bound
also. We then evaluate our upper bound and show that it meets the
rate achievable with our proposed signalling scheme. In this 2-2-1 system, the fact that both in our
achievable scheme and in our upper bound, the optimal transmit
covariance matrices turn out to be unit-rank, proves to be crucial in
enabling us to characterize the lower and upper bounds explicitly and
showing that they are equal.

Secure communications in multi-user networks, e.g., multiple access
channel \cite{liang-poor, liu-yates, tekin-yener, tekin-yener-mac,
ruoheng-itw}, broadcast channel \cite{LiuPoor:2007}, relay channel \cite{lai-elgamal,oohama}, interference
channel \cite{marij-interference}, and two-way channel
\cite{tekin-yener-2way}, and in fading channels
\cite{barros-rodrigues, liang-poor-shamai, li-yates-trappe,
gopala-lai-elgamal, li-yates-trappe-fading, Shafiee:2007} have been
considered recently.

We use the following notations throughout this paper: Bold face lower
and upper case letters are used to represent vectors and matrices,
respectively. $\mathbf{x}^T$ and $|| \mathbf{x} ||$ denote the
transpose and the Euclidean norm of the vector $\mathbf{x}$,
respectively. $\tr(\mathbf{X})$ and $|\mathbf{X}|$ denote the trace
and the determinant of the square matrix $\mathbf X$,
respectively. Whether a variable is deterministic or random will be
clear from the context.

%================================================================
\section{System Model}
The 2-2-1 Gaussian MIMO wire-tap channel is characterized by
\begin{align}
\mathbf{y} &=\mathbf{H} \mathbf{x}+\mathbf{n}_{y} \label{original1}\\
z& =\mathbf{g}^T \mathbf{x} +n_{z} \label{original2}
\end{align}
where $\mathbf{x}$ is the transmitted signal, and $\mathbf{y}$, $z$
are the received signals at the legitimate user and the eavesdropper,
respectively.  $\mathbf{n}_{y}$ is a Gaussian random vector with
zero-mean and identity covariance matrix, while $n_{z}$ is a Gaussian
random variable with zero-mean and unit-variance. $\mathbf{n}_{y}$,
$n_{z}$ are assumed to be independent. The transmitted signal
satisfies an average power constraint,
\begin{align}
\frac{1}{n} \sum_{i=1}^n E \left[\mathbf{x}_i^T \mathbf{x}_i
\right] \leq P \label{averagepower}
\end{align}
The secrecy capacity $C(P)$ is defined as the maximum number of bits
that can be correctly transmitted to the intended receiver while the
eavesdropper is essentially no better informed about the transmitted
information after observing the received signal than it was before
\cite{Hellman:1978}.

When $\Hmat$ is not full-rank, by performing singular value
decomposition (SVD) on $\Hmat$ and obtaining an equivalent channel by
rotation, it can be shown that the system is equivalent to a 2-1-1
system, whose secrecy capacity has been found in
\cite{Khisti:ISIT, Khisti:MISOME}. Therefore, without loss of generality, for the
rest of the paper, we assume that $\Hmat$ is full-rank, and hence is
invertible. When
\begin{align}
\left\|\mathbf{H}^{-T}\mathbf{g} \right\| \leq 1 \label{degradedcon1}
\end{align}
$z$ can be written as a noisy version of $\mathbf{y}$, i.e.,
$\mathbf{\mathbf{r}}^T \mathbf{y}+n$, which means that the channel is degraded.
In this case, no pre-processing of information is necessary
\cite{Csiszar:1978}, and also it can be shown that Gaussian signalling is
optimal. Thus, in this paper, we concentrate on the more interesting
and difficult case where $\Hmat$ is full-rank and satisfies
\begin{align}
\left\|\mathbf{H}^{-T}\mathbf{g} \right\| >1 \label{degradedcon}
\end{align}

\section{An Achievable Scheme}
By \cite{Csiszar:1978}, the following secrecy rate is achievable,
\begin{align}
\left[I(\mathbf{u};\mathbf{y})-I(\mathbf{u};z)\right]^+ \label{ck}
\end{align}
where $\mathbf{u} \rightarrow \mathbf{x} \rightarrow \mathbf{y}z$. By
taking $\mathbf{u}=\mathbf{x}$ and constraining the input signal
$\mathbf{x}$ to be Gaussian with covariance matrix $\mathbf{S}$ such
that $\tr(\mathbf{S}) \leq P$, the following secrecy rate is
achievable,
\begin{align}
\left[ \frac{1}{2}\log \left|\mathbf{I}+\mathbf{H} \mathbf{S} 
\mathbf{H}^T \right|-\frac{1}{2} \log (1+
\mathbf{g}^T \mathbf{S} \mathbf{g})\right]^+ \label{achieveMISO}
\end{align}
Thus, the following secrecy rate is achievable
\begin{align}
\max_{\mathbf{S} \succeq \mathbf{0}: \tr(\mathbf{S}) \leq P} \quad
\frac{1}{2} \log \left|\mathbf{I}+\mathbf{H} \mathbf{S} \mathbf{H}^T \right|- \frac{1}{2} \log (1+
\mathbf{g}^T \mathbf{S} \mathbf{g})  \label{ISIT}
\end{align}
unless the maximum value in (\ref{ISIT}) is negative, in which case,
the achieved secrecy rate is zero.

Ignoring the $1/2$, we may rewrite the cost function in (\ref{ISIT}) as
\begin{align}
\log \left|\Imat+\Hmat \Smat \Hmat^T \right|- \log \left(1+\gv^T \Smat \gv \right)=\log  \left|\Imat+ \Hmat^T \Hmat \Smat  \right|- \log \left(1+ \gv^T \Smat \gv \right) \label{similar0}
\end{align}
We first use the following lemma to show that the $\mathbf{S}$ that
maximizes (\ref{ISIT}) is unit-rank.
\begin{lemma} \label{KKTlemma}
If $\Dmat$ is a $2 \times 2$ invertible matrix that satisfies
\begin{align}
\gv^T \Dmat^{-1} \gv \geq 1
\end{align}
then the optimal $\Smat$ that solves the following optimization problem
\begin{align}
\max_{\Smat \succeq \mathbf{0}, \mathrm{tr}(\Smat) \leq P} \quad \log  \left|\Imat+ \Dmat \Smat  \right|- \log \left(1+ \gv^T \Smat \gv \right)  \label{optpro}
\end{align}
is unit-rank.
\end{lemma}

\begin{proof}
The KKT necessary conditions for the optimization problem in
(\ref{optpro}) are
\begin{align}
\Smat^* &\succeq \mathbf{0}\\
\tr(\Smat^*) &\leq P\\
\Cmat &\succeq \mathbf{0}\\
\lambda &\geq 0\\
\lambda (\tr(\Smat^*)-P) &=0\\
\Cmat \Smat^*&=\mathbf{0} \label{000}\\
- (\Imat+ \Dmat \Smat^* )^{-1} \Dmat +\frac{1}{1+\gv^T \Smat^* \gv}\gv \gv^T-\Cmat+\lambda \Imat&=\mathbf{0} \label{mainKKT}
\end{align}
We will prove the claim by contradiction. Assume that the optimal
$\Smat$ is full-rank. Then, from (\ref{000}), it follows that
$\Cmat=\mathbf{0}$, i.e, (\ref{mainKKT}) becomes
\begin{align}
(\Imat+\Dmat \Smat^* )^{-1} \Dmat =\frac{1}{1+\gv^T \Smat^* \gv}\gv \gv^T+\lambda \Imat
\end{align}
Since $\Dmat$ is invertible
\begin{align}
(\Imat+ \Dmat \Smat^* )^{-1}=\frac{1}{1+\gv^T \Smat^* \gv} \gv \gv^T \Dmat^{-1}+\lambda \Dmat^{-1}
\end{align}
Using the matrix inversion lemma \cite[page 19]{horn-johnson:book}, we have
\begin{align}
\Imat+\Dmat \Smat^* =\frac{1}{\lambda} \Dmat-\frac{1}{\lambda^2+\lambda^2 \gv^T \Smat^* \gv+\lambda ||\gv||^2} \Dmat \gv \gv^T
\end{align}
i.e.,
\begin{align}
\Smat^*=\frac{1}{\lambda} \Imat-\frac{1}{\lambda^2+\lambda^2 \gv^T \Smat^* \gv+\lambda ||\gv||^2}\gv \gv^T-\Dmat^{-1} \label{above}
\end{align}
We multiply both sides of (\ref{above}) with $\gv^T$ on the left and $\gv$ on the
right. Let us define $\gamma=\gv^T \Smat^* \gv$, which is a
non-negative real number. Then, we have
\begin{align}
\gamma=\frac{||\gv||^2}{\lambda}-\frac{||\gv||^4}{\lambda^2+\lambda^2 \gamma+\lambda ||\gv||^2}-\gv^T \Dmat^{-1} \gv
\end{align}
i.e., we have
\begin{align}
\gamma^2+\left(1+\gv^T \Dmat^{-1} \gv\right)\gamma+\gv^T \Dmat^{-1} \gv+\frac{||\gv||^2}{\lambda} \left(\gv^T \Dmat^{-1} \gv-1 \right)=0 \label{contra1}
\end{align}
Because $\gv^T \Dmat^{-1} \gv-1 \geq 0$, the second-order equation in
(\ref{contra1}) has no non-negative roots, i.e., it either has no real
roots, or it has two negative roots. Thus, we arrive at a
contradiction. Therefore, $\Cmat$ cannot be equal to $\mathbf{0}$, and
consequently, $\Smat$ cannot be full-rank, and it has to be unit-rank.
\end{proof}

Since $\Hmat^T \Hmat$ is invertible and satisfies (\ref{degradedcon}),
$\Dmat=\Hmat^T \Hmat$ satisfies the condition of Lemma
\ref{KKTlemma}. Hence, the optimal $\Smat$ for the optimization problem
in (\ref{ISIT}) is unit-rank.

Given that the optimal $\mathbf{S}$ is unit-rank, it can be written as
\begin{align}
\mathbf{S}=P \mathbf{q} \mathbf{q}^{T}
\end{align}
The corresponding achievable secrecy rate is
\begin{align}
R&=\frac{1}{2} \log \left|\mathbf{I}+P \mathbf{H} \mathbf{q} \mathbf{q}^{T} \mathbf{H}^T\right|- \frac{1}{2} \log (1+
P \mathbf{g}^T  \mathbf{q} \mathbf{q}^{T} \mathbf{g})\\
&=\frac{1}{2} \log \frac{\mathbf{q}^{T} (\mathbf{I} + P \mathbf{H}^T \mathbf{H}) \mathbf{q}}{\mathbf{q}^{T} (\mathbf{I} + P \mathbf{g} \mathbf{g}^T) \mathbf{q}}\label{rayleigh}
\end{align}
where (\ref{rayleigh}) is now in the Rayleigh quotient \cite[page
176]{horn-johnson:book} form and the optimal achievable $\mathbf{q}$, which we will
call $\qv_a$, is
\begin{align}
\mathbf{q}_a=\frac{\mathbf{B}^{-1/2}
\mathbf{w}_a}{||\mathbf{B}^{-1/2} \mathbf{w}_a||} \label{defineqa}
\end{align}
where $\mathbf{w}_a$ is the eigenvector that corresponds to the
largest eigenvalue of $\mathbf{B}^{-1/2} \mathbf{A} \mathbf{B}^{-1/2}$
with
\begin{align}
\mathbf{A}= & \mathbf{I}+P \mathbf{H}^T \mathbf{H}\\
\mathbf{B}= & \mathbf{I}+P \mathbf{g} \mathbf{g}^T
\end{align}
In other words, $\qv_a$ is the unit-norm eigenvector that satisfies
\begin{align}
\left( \mathbf{I}+P \mathbf{g} \mathbf{g}^T \right)^{-1} \left( \mathbf{I}+P \mathbf{H}^T \mathbf{H} \right) \qv_a=\lambda_1 \qv_a \label{propqa}
\end{align}
where $\lambda_1$ is the largest eigenvalue of the matrix
\begin{align}
\left( \mathbf{I}+P \mathbf{g} \mathbf{g}^T \right)^{-1/2} \left( \mathbf{I}+P \mathbf{H}^T \mathbf{H} \right) \left( \mathbf{I}+P \mathbf{g} \mathbf{g}^T \right)^{-1/2}
\end{align}
Written explicitly, the achievable secrecy rate is
\begin{align}
\frac{1}{2} \log \left(\frac{1+P \mathbf{q}_a^{T} \mathbf{H}^T
\mathbf{H} \mathbf{q}_a}{1+P \mathbf{q}_a^T \mathbf{g}
\mathbf{g}^T \mathbf{q}_a}\right)=\frac{1}{2} \log \lambda_1 \label{achievability}
\end{align}

Next, we show that the secrecy rate in (\ref{achievability}) is in
fact strictly positive. By picking $\Smat=P \gv^\perp \left(
\gv^\perp \right)^T$, where $\gv^\perp$ is the unit-norm vector that
is orthogonal to $\gv$, an achievable secrecy rate is
\begin{align}
\frac{1}{2} \log\left(1+P \left\|\Hmat \gv^\perp \right\|^2 \right) \label{gorth}
\end{align}
Since $\Hmat$ is full rank, $\Hmat \gv^\perp \neq \mathbf{0}$, i.e.,
the secrecy rate in (\ref{gorth}) is strictly positive. Since the
secrecy rate in (\ref{achievability}) is the maximum over all $\Smat$
satisfying $\tr(\Smat) \leq P$, we conclude that
\begin{align}
\frac{1}{2} \log \lambda_1 \geq \frac{1}{2} \log\left(1+P \left\|\Hmat \gv^\perp \right\|^2 \right) >0
\end{align}
which also means that
\begin{align}
\lambda_1>1 \label{lambdag1}
\end{align}

\section{A Tight Upper Bound}
The following theorem provides an upper bound on the secrecy capacity of the wire-tap channel described in (\ref{original1}) and (\ref{original2}).

\begin{theorem} \label{converse}
An upper bound on the secrecy capacity of the wire-tap channel
described in (\ref{original1}) and (\ref{original2}) is
\begin{align}
\max_{\Smat \succeq \mathbf{0}, \mathrm{tr}(\Smat) \leq P} U(\Smat, \av) 
\label{upper111}
\end{align}
for any $\av$ with $||\av|| < 1$, where $U(\Smat, \av)$ is defined
as
\begin{align}
U(\mathbf{S},\av) &= \frac{1}{2}\log
\frac{\left|\Imat+\Nmat^{-1}\bar{\Hmat} \Smat \bar{\Hmat}^T \right|}{\left(1+ \gv^T \Smat \gv \right) } \label{defineUSA}
\end{align}
with $\Nmat$ defined as
\begin{align}
\Nmat=\begin{bmatrix} \Imat & \av \\ \av^T & 1 \end{bmatrix} \label{defineNmat}
\end{align}
and $\bar{\Hmat}$ defined as
\begin{align}
\bar{\Hmat}=\begin{bmatrix}\Hmat \\ \gv^T \end{bmatrix} \label{defineHbar}
\end{align}
\end{theorem}

The proof of Theorem~\ref{converse} is provided in the Appendix. Intuitively, this upper bound is obtained by considering the secrecy capacity of a new channel where the legitimate receiver also has access to the eavesdropper's signal. Since the legitimate user is more capable in the new channel, the secrecy capacity of the new channel will serve as an upper bound on the secrecy capacity of the original channel. The new channel is degraded, and therefore the secrecy capacity is easier to obtain.

The vector $\av$ introduced in
Theorem \ref{converse} is the correlation between the Gaussian noises
at the legitimate user and the eavesdropper, i.e.,
\begin{align}
\av=E[\mathbf{n}_y n_z]
\end{align}
We note that $\av$ thus defined has to satisfy $||\av|| \leq 1$ for
$\Nmat$ in (\ref{defineNmat}) to be positive semi-definite. Introducing
correlation between $\mathbf{n}_y$ and $n_z$ does not change the
secrecy capacity of the channel, but changes the upper bound in
(\ref{upper111}). In fact, (\ref{upper111}) remains a valid upper bound for any $\av$, with $||\av||<1$. Thus, we will smartly pick an $\av$ vector, and show
that the upper bound with this $\av$ vector is in fact tight, to establish the secrecy capacity.

We rewrite $U(\mathbf{S},\av)$ as
\begin{align}
U(\mathbf{S},\av) 
%& =\frac{1}{2}\log
%\frac{\left|\Imat +\Nmat^{-1}\bar{\Hmat} \Smat \bar{\Hmat}^T \right|}{\left(1+ \gv^T \Smat \gv \right) }\\
& =\frac{1}{2}\log
\frac{\left|\Imat +\bar{\Hmat}^T \Nmat^{-1} \bar{\Hmat} \Smat  \right|}{\left(1+ \gv^T \Smat \gv \right) } \label{similar1}
\end{align}
By the definition of $\Nmat$ in (\ref{defineNmat}), we have
\begin{align}
\Nmat^{-1}=\begin{bmatrix} \Imat+\frac{1}{k} \av \av^T & -\frac{1}{k} \av \\
-\frac{1}{k} \av^T & \frac{1}{k} \end{bmatrix}
\end{align}
where $k=1-||\av||^2$. Then,
\begin{align}
\bar{\Hmat}^T \Nmat^{-1} \bar{\Hmat}&=\Hmat^T \Hmat+\frac{1}{k}\Hmat^T \av \av^T \Hmat-\frac{1}{k} \gv \av^T \Hmat-\frac{1}{k} \Hmat^T \av \gv^T+\frac{1}{k} \gv \gv^T\\
&=\Hmat^T \Hmat+\frac{1}{k} \left(\Hmat^T \av-\gv \right) \left(\Hmat^T \av-\gv \right)^T
\end{align}
Let us define $\Amat(\av)$ as
\begin{align}
\label{defineAmata}
\Amat(\av)=\bar{\Hmat}^T \Nmat^{-1} \bar{\Hmat}=\Hmat^T \Hmat+\frac{1}{k} \left(\Hmat^T \av-\gv \right) \left(\Hmat^T \av-\gv \right)^T
\end{align}
Then, $U(\Smat, \av)$ in (\ref{similar1}) is written as
\begin{align}
U(\Smat, \av)=\frac{1}{2} \log \left|\Imat+\Amat (\av) \Smat \right|-\frac{1}{2} \log \left(1+ \gv^T \Smat \gv \right) \label{ulukusadd}
\end{align}
Let us also define $\qv_a^\perp$ to be the unit-norm vector that is orthogonal to $\qv_a$, which is defined in (\ref{defineqa}).

We pick $\av$ to be of the form
\begin{align}
\av=\Hmat^{-T} \left(\alpha
\qv_a^\perp+\gv\right) \label{form}
\end{align}
for any real number $\alpha$ that makes $||\av||<1$. $\alpha=0$ results in $\av=\Hmat^{-T} \gv$, which is a vector with norm greater than 1, and therefore, is not permissible.

Then, with this selection of $\av$, $\Amat(\av)$ in
(\ref{defineAmata}) can be written as
\begin{align}
\Amat(\av)=\Hmat^T \Hmat+\theta(\alpha) \qv_a^\perp \left(\qv_a^\perp \right)^T \label{Aa}
\end{align}
where $\theta(\alpha)$ is defined as
\begin{align}
\theta(\alpha)& =\frac{\alpha^2}{1-\av^T \av} \label{definethetaalpha}\\
&=\frac{\alpha^2}{1-\left(\Hmat^{-T} \left(\alpha
\qv_a^\perp+\gv\right) \right)^T \left(\Hmat^{-T} \left(\alpha
\qv_a^\perp+\gv\right) \right)}
\end{align}
Then, we have
\begin{align}
\frac{1}{\theta(\alpha)}=-\left(\qv_a^\perp \right)^T \left(\Hmat^T \Hmat \right)^{-1} \qv_a^\perp-\frac{2 \gv^T \left(\Hmat^T \Hmat \right)^{-1} \qv_a^\perp}{\alpha}-\frac{\gv^T \left(\Hmat^T \Hmat \right)^{-1} \gv-1}{\alpha^2} \label{thetatheta}
\end{align}
This is a second-order polynomial in terms of $1/\alpha$, and it is
easy to see that $1/\alpha^*$ maximizes $\theta(\alpha)$, with
\begin{align}
\frac{1}{\alpha^*}=\frac{\gv^T \left(\Hmat^T \Hmat \right)^{-1} \qv_a^\perp}{1-\gv^T \left(\Hmat^T \Hmat \right)^{-1} \gv} \label{alphaalpha}
\end{align}
Finally, we call the $\av$ vector that we pick $\av^*$, which 
is given as
\begin{align}
\av^*=\Hmat^{-T} \left(\alpha^*
\qv_a^\perp+\gv\right)
\end{align}

First, we will prove that $\av^*$ has norm no greater than 1. Let us define
$\av_0$ to be
\begin{align}
\av_0=\frac{\gv^T \qv_a}{||\Hmat \qv_a||^2} \Hmat \qv_a
\end{align}
$\av_0$ satisfies the form of $\av$ in (\ref{form}) because $\Hmat^T
\av_0-\gv$ is orthogonal to $\qv_a$, hence, it is along the direction
of $\qv_a^\perp$. Therefore, $\av_0$ must correspond to an $\alpha$,
which we call $\alpha_0$. It can be seen that
\begin{align}
||\av_0||=\frac{\left|\gv^T \qv_a\right|}{||\Hmat \qv_a||} < 1 \label{normless}
\end{align}
because of (\ref{lambdag1}) and the fact that $\qv_a$ satisfies (\ref{achievability}), i.e.,
\begin{align}
1<\lambda_1=\frac{1+P ||\Hmat \qv_a||^2}{1+P (\gv^T \qv_a)^2}
\end{align}
Hence, (\ref{normless}) means that $\alpha_0 \neq 0$ and furthermore, we have
\begin{align}
\theta(\alpha_0)=\frac{\alpha_0^2}{1-\av_0^T \av_0}>0
\end{align}
Therefore, we have
\begin{align}
\frac{1}{\theta(\alpha^*)} \overset{(a)}{\geq} \frac{1}{\theta(\alpha_0)} > 0 \label{theta0}
\end{align}
where $(a)$ follows because $\alpha^*$ maximizes $\frac{1}{\theta(\alpha^*)}$. Finally, (\ref{theta0}) implies $||\av^*||<1$ because of (\ref{definethetaalpha}).

Next, we will show that the optimal $\Smat$ for $\max U(\Smat,\av^*)$ in (\ref{upper111}) is
unit-rank. Since the upper bound and achievable scheme differ only in
replacing $\Amat(\av^*)$ with $\Hmat^T \Hmat$, as shown in
(\ref{ISIT}) and (\ref{ulukusadd}), we will use Lemma
\ref{KKTlemma} again to show the optimality of unit-rank $\Smat$ in
(\ref{upper111}). Since $\Hmat^T \Hmat$ is invertible and
$\theta(\alpha^*)>0$, matrix $\Amat(\av^*)$, in the form of
(\ref{Aa}), is invertible. 
%Next, we will show that $\gv^T
%\Amat(\av^*)^{-1} \gv=1$. 
In addition, in order to use Lemma 1, we need $\gv^T \Amat(\av^*)^{-1}\gv \geq 1$. In
the
following, we will show that $\gv^T \Amat(\av^*)^{-1}\gv = 1$.
Using the matrix inversion lemma \cite[page
19]{horn-johnson:book} on (\ref{Aa}), we have
\begin{align}
\Amat(\av^*)^{-1}=\left(\Hmat^T \Hmat \right)^{-1}-\frac{1}{\frac{1}{\theta(\alpha^*)}+ \left(\qv_a^\perp \right)^T \left(\Hmat^T \Hmat \right)^{-1} \qv_a^\perp} \left(\Hmat^T \Hmat \right)^{-1} \qv_a^\perp \left(\qv_a^\perp \right)^T \left(\Hmat^T \Hmat \right)^{-1} \label{inverse}
\end{align}
Also, from (\ref{thetatheta}) and (\ref{alphaalpha}),  $1/\theta(\alpha^*)$ is equal to
\begin{align}
\frac{1}{\theta(\alpha^*)}&=-\left(\qv_a^\perp \right)^T \left( \Hmat^T \Hmat \right)^{-1} \qv_a^\perp+\frac{\left( \gv^T \left( \Hmat^T \Hmat \right)^{-1} \qv_a^\perp \right)^2 }{\gv^T \left( \Hmat^T \Hmat \right)^{-1} \gv-1 } \label{plug2}\\
&=-\left(\qv_a^\perp \right)^T \left( \left( \Hmat^T \Hmat \right)^{-1}-\frac{\left( \Hmat^T \Hmat \right)^{-1} \gv \gv^T \left( \Hmat^T \Hmat \right)^{-1}}{\gv^T \left( \Hmat^T \Hmat \right)^{-1} \gv-1}\right) \qv_a^\perp\\
&=-\left(\qv_a^\perp \right)^T \left(\Hmat^T \Hmat-\gv \gv^T \right)^{-1} \qv_a^\perp \label{oneover}
\end{align}
Now, using straightforward algebra, starting from (\ref{inverse}) and
(\ref{plug2}), it is easy to verify that
\begin{align}
\gv^T \Amat(\av^*)^{-1} \gv=1
\end{align}
Thus, $\Dmat=\Amat(\av^*)$ satisfies the conditions of Lemma
\ref{KKTlemma}, and therefore, $\arg \max U (\Smat, \av^*)$ is unit-rank.

Thus, for the selected $\av^*$, the optimization in the upper bound
in (\ref{upper111}) over $\Smat \succeq \mathbf{0}$ reduces to
an optimization over $\qv$, as $\Smat=P \qv \qv^T$, 
\begin{align}
\max_{\Smat \succeq \mathbf{0},  \tr(\Smat) \leq P} \quad U(\Smat, \av^*)=\max_{\qv} \quad \frac{1}{2} \log  \frac{\qv^T \left(\Imat+P \Hmat^T \Hmat+ P \theta(\alpha^*) \qv_a^\perp \left(\qv_a^\perp \right)^T \right) \qv}{\qv^T \left(\Imat+ P \gv \gv^T\right) \qv}  \label{problem}
\end{align}
where (\ref{problem}) is again in the Rayleigh quotient \cite[page 176]{horn-johnson:book}
form, and the solution to this optimization problem is the largest
eigenvalue of the matrix
\begin{align}
(\Imat+P \gv \gv^T)^{-1/2} \left(\Imat+P \Hmat^T \Hmat+ P \theta(\alpha^*) \qv_a^\perp \left(\qv_a^\perp \right)^T \right) (\Imat+P \gv \gv^T)^{-1/2} \label{symmetric}
\end{align}
which is the largest eigenvalue of the matrix
\begin{align}
(\Imat+P \gv \gv^T)^{-1} \left(\Imat+P \Hmat^T \Hmat+ P \theta(\alpha^*) \qv_a^\perp \left(\qv_a^\perp \right)^T \right) \label{nonsymmetric}
\end{align}
since the two matrices are related by a similarity transformation.
Note that
\begin{align}
(\Imat+P \gv \gv^T)^{-1} \left(\Imat+P \Hmat^T \Hmat+ P \theta(\alpha^*) \qv_a^\perp \left(\qv_a^\perp \right)^T \right) \qv_a&=(\Imat+P \gv \gv^T)^{-1} \left(\Imat+P \Hmat^T \Hmat \right) \qv_a \\
&=\lambda_1 \qv_a \label{eigen}
\end{align}
where (\ref{eigen}) follows from (\ref{propqa}).

Let us define vector $\qv_1$ as
\begin{align}
\qv_1=- \theta(\alpha^*) \left(\Hmat^T \Hmat-\gv \gv^T \right)^{-1} \qv_a^\perp \label{defineq1}
\end{align}
Note that
\begin{align}
\qv_1^T \qv_a^\perp=-\theta(\alpha^*) \left(\qv_a^\perp \right)^T\left(\Hmat^T \Hmat-\gv \gv^T \right)^{-1} \qv_a^\perp=1 \label{111}
\end{align}
where the last equality follows from (\ref{oneover}).
Also, (\ref{defineq1}) implies that
\begin{align}
\Hmat^T \Hmat \qv_1=\gv \gv^T \qv_1-\theta(\alpha^*) \qv_a^\perp \label{Ulukus2}
\end{align}
Then, we have
\begin{align}
&(\Imat+P \gv \gv^T)^{-1} \left(\Imat+P \Hmat^T \Hmat+ P \theta(\alpha^*) \qv_a^\perp \left(\qv_a^\perp \right)^T \right) \qv_1 \nonumber \\
& \hspace{0.3in}=(\Imat+P \gv \gv^T)^{-1} \left(\left(\Imat+P \Hmat^T \Hmat \right) \qv_1+P \theta(\alpha^*) \qv_a^\perp \right) \label{q1qa1}\\
& \hspace{0.3in}= (\Imat+P \gv \gv^T)^{-1} \left(\qv_1+P\gv \gv^T \qv_1-P\theta(\alpha^*) \qv_a^\perp+P \theta(\alpha^*) \qv_a^\perp \right) \label{Ulukus}\\
& \hspace{0.3in}= (\Imat+P \gv \gv^T)^{-1} (\Imat+P \gv \gv^T) \qv_1\\
&\hspace{0.3in}= \qv_1
\end{align}
where (\ref{q1qa1}) follows from (\ref{111}), and (\ref{Ulukus})
follows from (\ref{Ulukus2}). This means that the eigenvalues of the
matrix in (\ref{nonsymmetric}), and also the eigenvalues of the matrix
in (\ref{symmetric}), are $\lambda_1$ and $1$. Since $\lambda_1>1$, as
shown in (\ref{lambdag1}), the resulting maximum value in
(\ref{problem}) is $\frac{1}{2} \log \lambda_1$. Hence, the upper
bound on the secrecy capacity, i.e.,$ \max_{\Smat \succeq \mathbf{0},
\tr(\Smat) \leq P} U(\Smat, \av^*)$, is $\frac{1}{2}\log \lambda_1$,
which is equal to the lower bound on the secrecy capacity shown in
(\ref{achievability}).

\section{Conclusions}

We determined the secrecy capacity of the 2-2-1 Gaussian MIMO wire-tap
channel, by solving for the optimum joint distribution for the
auxiliary random variable and the channel input in the Csiszar-Korner
formula. First, we proposed a lower bound on the secrecy capacity by
evaluating the Csiszar-Korner formula for a specific selection of the
auxiliary random variable and the channel input. Our achievable scheme
is based on Gaussian signalling and no pre-processing of information.
Even for this achievable scheme, which is completely characterized by
the transmit covariance matrix $\Smat$, a closed form solution for the
secrecy rate does not exist. However, in our 2-2-1 case, we have shown
that the optimal transmission scheme is unit-rank, i.e., beam-forming
is optimal.

We showed the optimality of the proposed achievable scheme by
constructing a tight upper bound that meets it. The upper bound is developed by considering the secrecy capacity of a channel where the
eavesdropper's signal is given to the legitimate receiver. Even though this upper
bound is well-defined for a general MIMO wire-tap channel, explicit
evaluation and tightening of this upper bound has been possible by
restricting ourselves to the 2-2-1 case. As in the lower-bound, and by
selecting a certain correlation structure for the additive noises, we
have shown that beam-forming is optimal for the upper bound as
well. Furthermore, we have shown that the optimal beam-forming
directions in the lower and upper bounds are the same. Finally, we
have shown that the two bounds meet yielding the secrecy capacity.

Our derivation is specific to the 2-2-1 case and we have not been able
to show that these lower and upper bounds meet in the general MIMO
channel. This is because the unit-rank (beam-forming) property of the
optimum transmit matrices is essential in our derivations, while
beam-forming is not likely to be the optimal strategy when the number
of transmit and receive antennas is more than two. Even though the
results presented in this paper are a non-trivial step towards the
solution of the general MIMO wire-tap problem, whether our techniques
can be useful in determining the secrecy capacity of larger MIMO
systems is unclear.

%-------------------------------------------------------------------------------------------------------------------------------------------
\section{Appendix}
\textbf{Proof of Theorem \ref{converse}}:
A proof of similar results is presented for the case of $m$-1-$n$ system, $m,n \geq 1$, in \cite[Lemma 1, 2]{Khisti:MISOME}. 
Our proof utilizes \cite[Lemma 1]{Khisti:MISOME}, which generalizes to the case of multiple antennas at the legitimate receiver easily, and extends \cite[Lemma 2]{Khisti:MISOME} to the case where there are two antennas at the legitimate receiver.

An upper bound on the secrecy capacity of the wire-tap channel described in (\ref{original1}) and (\ref{original2}) is \cite[Lemma 1]{Khisti:MISOME}
\begin{align}
\max_{p(\mathbf{x}):E[\mathbf{x}^T \mathbf{x}] \leq P} \quad I(\mathbf{x};\mathbf{y}|z) \label{Wornell}
\end{align}
Since we have
\begin{align}
I(\mathbf{x};\mathbf{y}|z)=I(\mathbf{x};\mathbf{y},z)-I(\mathbf{x};z) \label{formula}
\end{align}
Intuitively, the upper bound is obtained by considering the secrecy capacity of a new channel where the legitimate receiver also has access to the eavesdropper's signal. Since the legitimate user is more capable in the new channel, the secrecy capacity of the new channel will serve as an upper bound on the secrecy capacity of the original channel. The new channel is degraded, and therefore the secrecy capacity formula is (\ref{formula}), obtained by setting $\mathbf{u}=\mathbf{x}$ as shown in \cite{Csiszar:1978}.

In evaluating the right-hand side of (\ref{Wornell}), we introduce correlation between $\mathbf{n}_y$ and $n_z$, i.e.,
let us define $\av$ to be 
\begin{align}
\av=E[\mathbf{n}_y n_z] \label{defineav}
\end{align}
We note that $\av$ thus defined has to satisfy $||\av|| \leq 1$. To avoid irregular cases, we will only consider $\av$ such that $||\av|| < 1$. We
also note that $\av$ does not affect the secrecy capacity of the original channel, but it affects the upper bound in (\ref{Wornell}). Thus, (\ref{Wornell}) remains an upper bound for
any $\av$ with $||\av|| < 1$.

We evaluate $I(\mathbf{x};\mathbf{y}|z)$ as follows,
\begin{align}
I(\mathbf{x};\mathbf{y}|z)&=h(\mathbf{y}|z)-h(\mathbf{y}|z,\mathbf{x}) \label{usestart}\\
&=h(\mathbf{y}|z)-h(\mathbf{n}_y|n_z)
\end{align}
Due to the Gaussianity of the noise,
\begin{align}
h(\mathbf{n}_y|n_z)=h(\mathbf{n}_y, n_z)-h(n_z)
=\frac{1}{2} \log (2 \pi e)^2 \left|\Nmat \right|
\end{align}
where $\Nmat$ is defined as in (\ref{defineNmat}).
Let us define $\Smat$  as
\begin{align}
\Smat=E[\mathbf{x} \mathbf{x}^T]
\end{align}
then 
\begin{align}
E[\mathbf{y} z]&=E\left[ \left(\Hmat \mathbf{x}+\mathbf{n}_y \right)\left(\mathbf{x}^T \gv+n_z \right) \right]=\Hmat \Smat \gv+\av\\
E[z^2]&=1+\gv^T \Smat \gv\\
E[\mathbf{y} \mathbf{y}^T]&=\Imat+\Hmat \Smat \Hmat^T
\end{align}
The linear minimum mean squared error (LMMSE) estimator of $\mathbf{y}$ using $z$ is
\begin{align}
\hat{\mathbf{y}}=\frac{\Hmat \Smat \gv+\av}{1+\gv^T \Smat \gv}z
\end{align}
and the resulting covariance matrix of the estimation error is
\begin{align}
\Imat+\Hmat \Smat \Hmat^T-\frac{1}{1+\gv^T \Smat \gv}\left(\Hmat \Smat \gv+\av \right)\left(\Hmat \Smat \gv+\av \right)^T
\end{align}
Hence,
\begin{align}
h(\mathbf{y}|z) &= h \left(\mathbf{y}-\frac{\Hmat \Smat \gv+\av}{1+\gv^T \Smat \gv}z\Big|z \right)\\
& \leq  h \left(\mathbf{y}-\frac{\Hmat \Smat \gv+\av}{1+\gv^T \Smat \gv}z\right)\\
&\leq \frac{1}{2} \log (2 \pi e)^2 \left|\Imat+\Hmat \Smat \Hmat^T-\frac{1}{1+\gv^T \Smat \gv}\left(\Hmat \Smat \gv+\av \right)\left(\Hmat \Smat \gv+\av \right)^T
 \right|
\end{align}
Therefore,  
\begin{align}
I(\mathbf{x};\mathbf{y}|z) & \leq \frac{1}{2} \log \frac{\left|\Imat+\Hmat \Smat \Hmat^T-\frac{1}{1+\gv^T \Smat \gv}\left(\Hmat \Smat \gv+\av \right)\left(\Hmat \Smat \gv+\av \right)^T
 \right|
}{\left|\Nmat \right|
}  \\
&=\frac{1}{2} \log \frac{\left|\left(\Imat+\Hmat \Smat \Hmat^T \right)\left(1+\gv^T \Smat \gv \right)-\left(\Hmat \Smat \gv+\av \right)\left(\Hmat \Smat \gv+\av \right)^T
 \right|
}{\left(1+\gv^T \Smat \gv \right)\left|\Nmat \right|
} \\
&=\frac{1}{2} \log \frac{\left|\begin{bmatrix} \Imat+\Hmat \Smat\Hmat^T & \Hmat \Smat \gv+\av\\
\gv^T \Smat \Hmat^T+\av^T & 1+\gv^T \Smat \gv \end{bmatrix}
 \right|
}{\left(1+\gv^T \Smat \gv \right)\left|\Nmat \right|
}  \\
&=\frac{1}{2} \log \frac{\left|\Nmat+\bar{\Hmat} \Smat \bar{\Hmat}^T \right|}{(1+\gv^T \Smat \gv) \left|\Nmat \right|}\\
&=\frac{1}{2} \log \frac{\left|\Imat+\Nmat^{-1}\bar{\Hmat} \Smat \bar{\Hmat}^T \right|}{(1+\gv^T \Smat \gv) } \label{usefinish2}
\end{align}
where $\bar{\Hmat}$ is defined as in (\ref{defineHbar}).
Thus, we have 
\begin{align}
\max_{p(\mathbf{x}):E[\mathbf{x}^T \mathbf{x}] \leq P} \quad I(\mathbf{x};\mathbf{y}|z) \leq  \max_{\Smat \succeq \mathbf{0}, \tr(\Smat) \leq P} \quad \frac{1}{2} \log \frac{\left|\Imat+\Nmat^{-1}\bar{\Hmat} \Smat \bar{\Hmat}^T \right|}{(1+\gv^T \Smat \gv) }  \label{usefinish}
\end{align}
Therefore, an upper bound on the secrecy capacity of the wire-tap channel described in (\ref{original1}) and (\ref{original2}) is
\begin{align}
\max_{\Smat \succeq \mathbf{0}, \tr(\Smat) \leq P} U(\Smat, \av)
\end{align}
for any $\av$ with $||\av|| < 1$, with $U(\Smat, \av)$ defined in (\ref{defineUSA}).

\bibliographystyle{unsrt}
\bibliography{ref}

%\begin{chapthebibliography}{1}

%\bibitem{e1} Bib Entry

%\end{chapthebibliography}

\end{document}